\documentstyle[epsfig,longtable,times]{mn}

\newif\ifAMStwofonts

\def\mabs{$M_{\rm B}$}

\def\hii{H{\sc ii}}

\def\doh{$12 + \log(\rm O/H)$}

\def\halpha{\ifmmode {\rm H{\alpha}} \else $\rm H{\alpha}$\fi}
\def\hbeta{\ifmmode {\rm H{\beta}} \else $\rm H{\beta}$\fi}
\def\oii{[O\,{\sc ii}] $\lambda$3727}
\def\oiii{[O\,{\sc iii}] $\lambda\lambda$4959,5007}
\def\oiiia{[O\,{\sc iii}] $\lambda$4959}
\def\oiiib{[O\,{\sc iii}] $\lambda$5007}
\def\oiiic{[O\,{\sc iii}] $\lambda$4363}
\def\nii{[N\,{\sc ii}] $\lambda$6584}

\def\sii{[S\,{\sc ii}] $\lambda\lambda$6717+6731}
\def\siii{[S\,{\sc ii}] $\lambda\lambda$6717,6731}
\def\rr23{$R_{\rm 23}$}
\def\oo32{$O_{\rm 32}$}

\ifoldfss
\newcommand{\rmn}[1] {{\rm #1}}

\renewcommand{\chem}[2] {$\rm{}^{#2}\kern-0.8pt#1$}
\ifCUPmtlplainloaded \else
  \NewTextAlphabet{textbfit} {cmbxti10} {}
  \NewTextAlphabet{textbfss} {cmssbx10} {}
  \NewMathAlphabet{mathbfit} {cmbxti10} {} 
  \NewMathAlphabet{mathbfss} {cmssbx10} {} 
\fi
\ifAMStwofonts
  \ifCUPmtlplainloaded \else
    \NewSymbolFont{upmath} {eurm10}
    \NewSymbolFont{AMSa} {msam10}
    \NewMathSymbol{\upi}     {0}{upmath}{19}
    \NewMathSymbol{\umu}     {0}{upmath}{16}
    \NewMathSymbol{\upartial}{0}{upmath}{40}
    \NewMathSymbol{\leqslant}{3}{AMSa}{36}
    \NewMathSymbol{\geqslant}{3}{AMSa}{3E}

    \let\leq=\leqslant \let\le=\leqslant
    \let\geq=\geqslant \let\ge=\geqslant
  \fi
\fi
\fi 

\ifnfssone
\newmathalphabet{\mathit}
\addtoversion{normal}{\mathit}{cmr}{m}{it}
\addtoversion{bold}{\mathit}{cmr}{bx}{it}
\newcommand{\rmn}[1] {\mathrm{#1}}

\newmathalphabet{\mathbfit} 
\addtoversion{normal}{\mathbfit}{cmr}{bx}{it}
\addtoversion{bold}{\mathbfit}{cmr}{bx}{it}
\newmathalphabet{\mathbfss} 
\addtoversion{normal}{\mathbfss}{cmss}{bx}{n}
\addtoversion{bold}{\mathbfss}{cmss}{bx}{n}
\ifAMStwofonts
  \ifCUPmtlplainloaded \else
    \UseAMStwoboldmath
    \makeatletter
    \new@mathgroup\upmath@group
    \define@mathgroup\mv@normal\upmath@group{eur}{m}{n}
    \define@mathgroup\mv@bold\upmath@group{eur}{b}{n}
    \edef\UPM{\hexnumber\upmath@group}
    \new@mathgroup\amsa@group
    \define@mathgroup\mv@normal\amsa@group{msa}{m}{n}
    \define@mathgroup\mv@bold\amsa@group{msa}{m}{n}
    \edef\AMSa{\hexnumber\amsa@group}
    \makeatother
    \mathchardef\upi="0\UPM19
    \mathchardef\umu="0\UPM16
    \mathchardef\upartial="0\UPM40
    \mathchardef\leqslant="3\AMSa36
    \mathchardef\geqslant="3\AMSa3E

    \let\leq=\leqslant \let\le=\leqslant
    \let\geq=\geqslant \let\ge=\geqslant
  \fi
\fi
\fi 

\ifnfsstwo
\newcommand{\rmn}[1] {\mathrm{#1}}

\DeclareMathAlphabet{\mathbfit}{OT1}{cmr}{bx}{it}
\SetMathAlphabet\mathbfit{bold}{OT1}{cmr}{bx}{it}
\DeclareMathAlphabet{\mathbfss}{OT1}{cmss}{bx}{n}
\SetMathAlphabet\mathbfss{bold}{OT1}{cmss}{bx}{n}
\ifAMStwofonts
  \ifCUPmtlplainloaded \else
    \DeclareSymbolFont{UPM}{U}{eur}{m}{n}
    \SetSymbolFont{UPM}{bold}{U}{eur}{b}{n}
    \DeclareSymbolFont{AMSa}{U}{msa}{m}{n}
    \DeclareMathSymbol{\upi}{0}{UPM}{"19}
    \DeclareMathSymbol{\umu}{0}{UPM}{"16}
    \DeclareMathSymbol{\upartial}{0}{UPM}{"40}
    \DeclareMathSymbol{\leqslant}{3}{AMSa}{"36}
    \DeclareMathSymbol{\geqslant}{3}{AMSa}{"3E}

    \let\leq=\leqslant \let\le=\leqslant
    \let\geq=\geqslant \let\ge=\geqslant
  \fi
\fi
\fi 

\ifCUPmtlplainloaded \else
\ifAMStwofonts \else 
  \def\upi{\pi}
  \def\umu{\mu}
  \def\upartial{\partial}
\fi
\fi

\title[L-Z relation in the local universe from the 2dFGRS]
{The Luminosity-Metallicity
relation in the local universe from the 2dF Galaxy Redshift Survey}
\author[F. Lamareille et al.]
{F. Lamareille,$^{1,2}$\thanks{E-mail:
flamare@ast.obs-mip.fr} M. Mouhcine,$^{3,4}$ T. Contini,$^{1}$
I. Lewis,$^{5}$ and S. Maddox$^{3}$\\
$^{1}$Laboratoire d'Astrophysique de Toulouse et Tarbes (UMR 5572), 
Observatoire Midi-Pyr\'en\'ees, 14 Avenue E. Belin, F-31400 Toulouse, France\\
$^{2}$Observatoire de Paris, 61 Avenue de l'Observatoire, F-75014 Paris,
France\\
$^{3}$School of Physics and Astronomy, University of Nottingham, University
Park, Nottingham NG7 2RD, UK\\
$^{4}$Observatoire Astronomique de Strasbourg (UMR 7550), 11 rue de 
l'Universit\'e, F-67000 Strasbourg, France\\
$^{5}$Department of Physics, Keble Road, Oxford OX1 3RH, UK}
\date{Accepted . Received ; in original form }

\pagerange{\pageref{firstpage}--\pageref{lastpage}} \pubyear{2004}

\begin{document}
\sloppy
\label{firstpage}

\maketitle

\begin{abstract}

We investigate the Luminosity -- Metallicity ($L-Z$) relation in the local 
universe ($0 < z < 0.15$) using spectra of 6\,387 star-forming galaxies 
extracted from the 2dF Galaxy Redshift Survey. This sample is by 
far the largest to date used to perform such a study. 
We distinguish star-forming galaxies from AGNs using ``standard'' diagnostic 
diagrams to build a homogeneous sample of starburst galaxies for the $L-Z$ 
study. We propose new diagnostic diagrams using ``blue'' emission lines (\oii, 
\oiiib, and \hbeta) only to discriminate starbursts from AGNs in intermediate-redshift 
($z > 0.3$) galaxies. 
Oxygen-to-hydrogen (O/H) abundance ratios are estimated using the "strong-line"
method, which relates the strength of following bright emission lines \oii, \oiiib, 
and \hbeta\ (parameters \rr23\ and \oo32) to O/H. We used the \nii/\halpha\ 
emission-line ratio as ``secondary'' abundance indicator to break the degeneracy 
between O/H and \rr23. 
We confirm the existence of the luminosity -- metallicity relation over a large 
range of abundances ($\sim 2$ dex) and luminosities ($\sim 9$ magnitudes). 
We find a linear relation between the gas-phase oxygen 
abundance and both the ``raw'' and extinction-corrected absolute $B$-band 
magnitude with a rms of $\sim 0.27$. A similar relation, with nearly the 
same scatter, is found in the $R$ band.
This relation is in good agreement with the one derived by Melbourne \& 
Salzer (2002) using the KISS data. However, our $L-Z$ relation is much 
steeper than previous determinations using samples of ``normal'' irregular 
and spiral galaxies. This difference seems to be primarily due to the 
choice of the galaxy sample used to investigate the $L-Z$ relation 
rather than any systematic error affecting the O/H determination. 
We anticipate that this luminosity -- metallicity relation will be used 
as the local ``reference'' for future studies of the evolution with 
cosmic time of fundamental galaxy scaling relations.  

\end{abstract}


\section{Introduction}

Our understanding of galaxy formation and evolution certainly benefits 
from improving our knowledge about the chemical properties of galaxies. 
The chemical composition of stars and gas within a galaxy depends on 
various physical processes, such as the star formation history, gas 
outflows and inflows, stellar initial mass function, etc. Although it 
is a complicated task to disentangle the effects of these various processes, 
the determination of galactic chemical abundances at various epochs put 
strong constraints on the likely evolutionary histories of galaxies (see 
Pettini 2003 for a review).

The correlation between galaxy metallicity and luminosity in the 
local universe is one of the most significant observational results 
in galaxy chemical evolution studies. Lequeux et al. (1979) first 
revealed that the oxygen abundance O/H increases with the total mass 
of irregular galaxies. To avoid several problems in the estimate of dynamical 
masses of galaxies, especially for irregulars, absolute magnitudes
are commonly used instead. The luminosity -- metallicity ($L-Z$) relation for 
irregulars was later confirmed by Skillman, Kennicutt \& Hodge (1989), 
Richer \& McCall (1995) and Pilyugin (2001a) among others. 
Subsequent studies have extended the $L-Z$ relation
to spiral galaxies (Garnett \& Shields 1987; Zaritsky et al. 1994; 
Garnett et al. 1997; Pilyugin \& Ferrini 2000), and to elliptical 
galaxies (Brodie \& Huchra 1991). The luminosity correlates with 
metallicity over $\sim 10$ magnitudes in luminosity and 2 dex in 
metallicity, with indications suggesting that the relationship may 
be environmental- (Vilchez 1995) and morphology- (Mateo 1998) free.
This suggests that similar phenomena govern the $L-Z$ over the 
whole Hubble sequence, from irregular/spirals to ellipticals 
(e.g. Garnett 2002; Pilyugin, Vilchez \& Contini 2004). 
Recently, Melbourne \& Salzer (2002) have used a sample of 519 star-forming 
emission-line galaxies from the KPNO International Spectroscopic Survey 
(KISS) to confirm the existence of the $L-Z$ relation over a broad range 
of luminosity and metallicity. They found however that the slope of the 
$L-Z$ relation is steeper than the dwarf galaxy $L-Z$ relation (see also 
Pilyugin \& Ferrini 2000). This may be evidence that the relationship is 
not linear over the full luminosity range (see also Mouhcine \& Contini 
2002). 

Many recent studies in galaxy evolution trace changes in scaling 
relations of galaxies as a function of cosmic epochs, such as the 
Tully -- Fisher relation for disks (e.g. Milvang-Jensen et al. 2003; 
Ziegler et al. 2002) and the fundamental plane relation for 
spheroids (e.g. van Dokkum \& Ellis 2003; Im et al. 2002). In this 
context, the luminosity -- metallicity relation of galaxies can be 
used as a sensitive probe and consistency check of galaxy evolution. 

The chemical properties of galaxies at different epochs provide new 
constraints on theories of galaxy formation and evolution. 
If local effects such as the gravitational potential and ``feedback'' 
from supernova-driven winds are the dominant regulatory mechanisms 
for star formation and chemical enrichment, then the $L-Z$ relation 
might be nearly independent of cosmic epoch, such as predicted by 
the semi-analytic models of galaxy formation and evolution (e.g. 
Kauffmann et al. 1993; Somerville \& Primack 1999). However, based 
on our current knowledge of galaxy evolution, suggesting that the 
cosmic star formation rate was higher in the past (e.g. Madau et al. 1996), 
and that the overall metallicity in the universe at earlier times 
was lower, we might expect galaxies to be considerably brighter at 
a given metallicity if there was more primordial gas available in 
the ``young'' universe to fuel star formation.

With the advent of the 10-m class telescopes and the powerful 
optical and near-infrared spectrographs, it is now possible to 
probe the physical properties (SFR, extinction, chemical abundances, 
mass, stellar populations, etc) of intermediate ($0 < z < 1$; 
Kobulnicky, Kennicutt \& Pizagno 1999; Hammer et al. 2001, Contini et al. 2002, 
Lilly, Carollo \&  Stockton 2003; Kobulnicky et al. 2003; Lamareille 
et al. in preparation) and high-redshift ($1.5 < z < 4$; Pettini et 
al. 1998, 2001; Kobulnicky \& Koo 2000; Mehlert et al. 2002; 
Lemoine-Busserolle et al. 2003; Erb et al. 2003) galaxies. Even if 
the number of galaxies per redshift bin is still small, these 
studies show that high-redshift galaxies ($z > 2$) are 2-4 magnitudes 
brighter than local galaxies of similar metallicity. 
This deviation from the local $L-Z$ relation demonstrates that the 
ratio between the luminosity and metal content varies throughout a 
galaxy's lifetime and is a powerful diagnostic of its evolutionary 
state. 

%

At intermediate redshifts, the situation is still not clear. 
Kobulnicky \& Zaritsky (1999) and Lilly, Carollo \&  Stockton (2003) 
found their samples of intermediate-$z$ galaxies to conform to 
the local $L-Z$ relation without any significant evolution of 
this relation out to $z \sim 1$. On the contrary, Kobulnicky 
et al. (2003) claimed recently that both the slope and zero point 
of the $L-Z$ relation evolve with redshift, the slope becoming 
steeper at early cosmic time. 

The determination of the fraction of galaxies that deviate from the 
local $L-Z$ relation, as well as the amplitude of this deviation as a 
function of redshift, will certainly help in constraining the ``differential'' 
rate of star formation between high-redshift galaxies and local ones. 
However, such studies require a reliable local $L-Z$ relation, extending 
over a large range of metallicities and luminosities, taking into account 
the variety of the local galaxy population and giving a better estimate 
of the overall shape of the relation and its intrinsic scatter. 

The main goal of this paper is thus to establish the luminosity -- 
metallicity relation in the local Universe. To do so we extracted a 
sub-sample of nearly 7\,000 emission-line star-forming galaxies 
with high signal-to-noise spectra from the 2dF Galaxy Redshift Survey 
(hereafter 2dFGRS). We anticipate that this luminosity -- metallicity 
relation will be used as the local {\it calibration} for future studies 
of the evolution with cosmic time of fundamental galaxy scaling relations.  

This paper is organized as follow. In Sect.~\ref{data}, we present the 
selection criteria applied to the original 2dFGRS sample to define 
a sub-sample of emission-line galaxies with high quality spectra. 
Various observed line ratios and their implications for the analysis 
of the physical conditions in emission-line galaxies, and the oxygen 
abundance estimates are also discussed. 
In Sect. \ref{lz} we present the derived luminosity -- metallicity relation, 
and show how it compares with previous determinations. In Sect. \ref{concl}, 
we present the results of this paper and summarize our conclusions. 

Throughout this paper, all calculations assume the cosmology given 
by \textit{WMAP}, with $\Omega_{\Lambda}=0.73$, $\Omega_{m}=0.27$ 
and $\rmn{H}_{0}=71\rmn{km}\ \rmn{s}^{-1}\ \rmn{Mpc}^{-1}$.


\begin{figure*}
\includegraphics[clip=,width=0.45\textwidth]{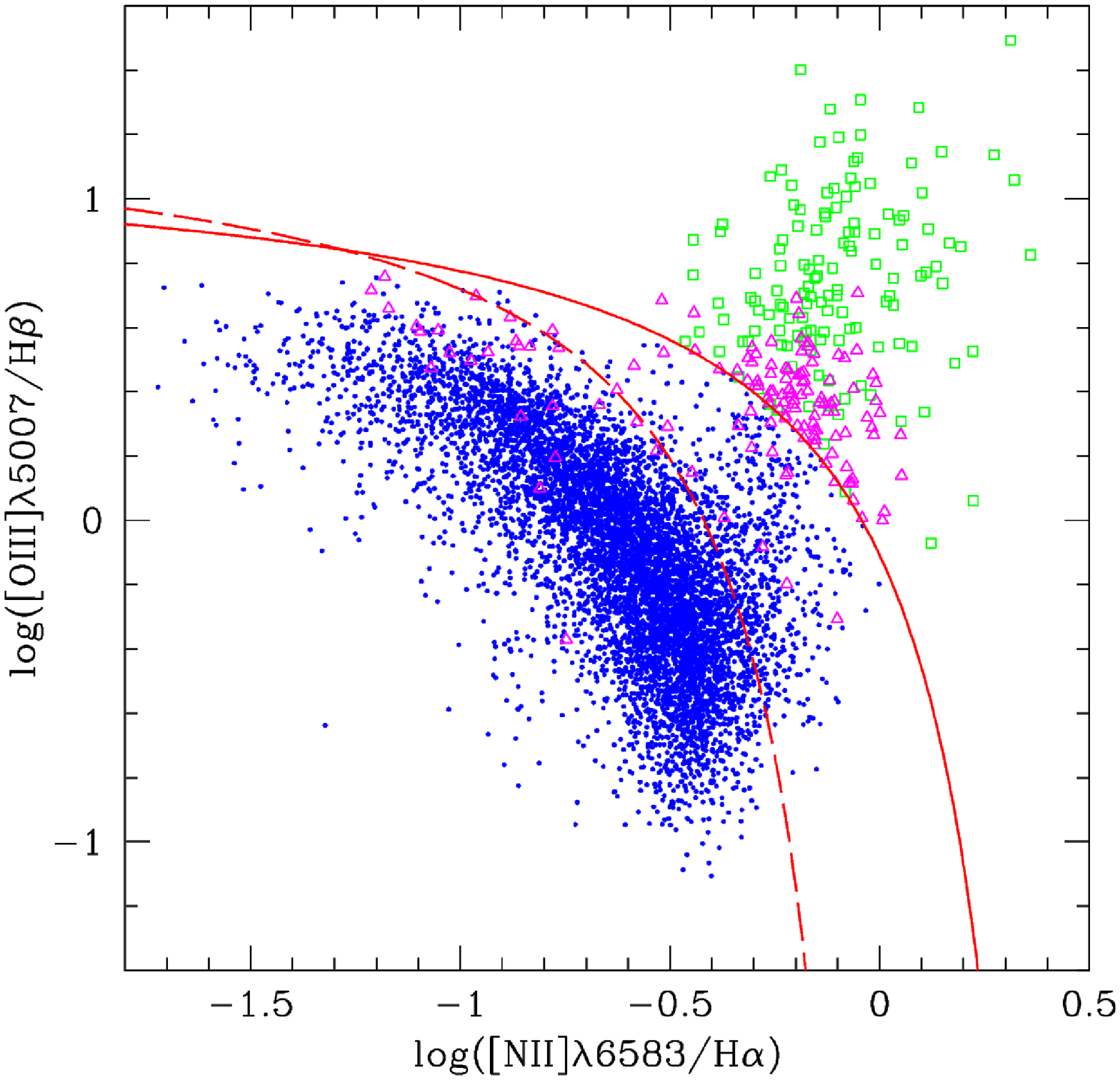}
\includegraphics[clip=,width=0.45\textwidth]{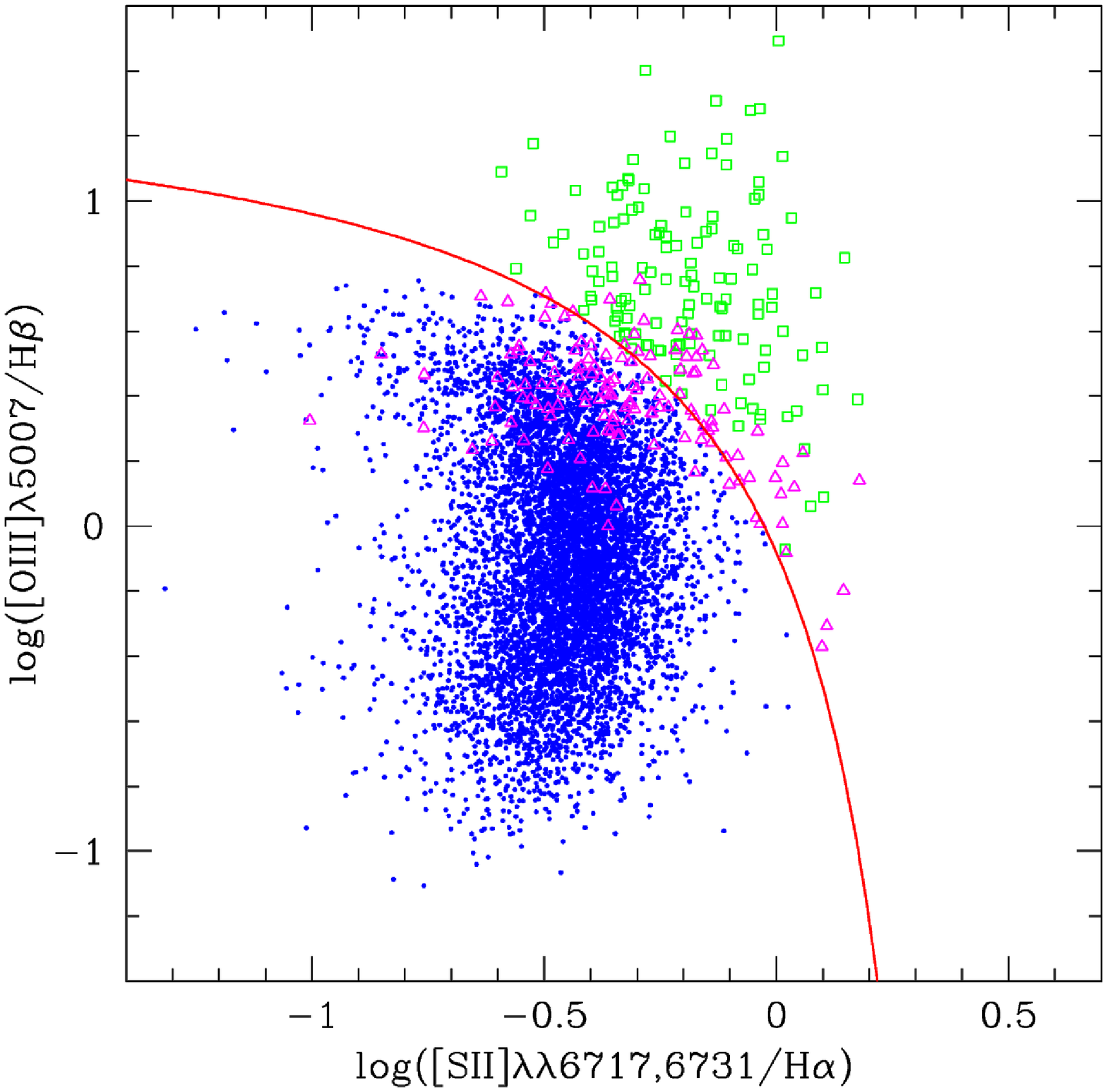}
\caption{Diagnostic diagrams for our sub-sample of 7\,353 narrow emission-line 
galaxies extracted from 2dFGRS. The continuous lines show the theoretical separation between 
starburst galaxies and AGNs from Kewley et al. (2001). An error of 0.15 dex 
has been added to the predicted separation in the \sii/\halpha\ diagram. 
The dashed line in the \nii/\halpha\ vs. \oiiib/\hbeta\ diagram shows the 
separation between starburst galaxies and AGNs as defined empirically by 
Kauffmann et al. (2003) using SDSS data. Dots represent star-forming galaxies, 
squares show AGNs, while triangles represent emission-line galaxies which 
are contradictory classified using both diagrams.}
\label{diag}
\end{figure*}

\section{The data}
\label{data}
\subsection{Sample selection}

The sample of emission-line star-forming galaxies used to establish the local 
$L-Z$ relation is extracted from the 2dFGRS observations.  

The 2dFGRS data set consists of optical ($3600-8000$\AA; $\Delta\lambda=10$\AA) 
spectroscopy of more than 250\,000 galaxies, covering two contiguous declination 
strips, plus 99 randomly located fields. One of the strips is located close to 
the super galactic plane, while the other strip is located in the northern 
galactic hemisphere. The mean redshift of the initial sample is $0.11$, 
with almost all galaxies with $z<0.3$. Full details of the survey strategy 
are given in Colless et al. (2001). Although, the original 2dFGRS spectra are 
not flux calibrated, the relative flux calibration, which is more critical for 
our analysis of emission-line ratios and equivalent widths is accurate 
(Lewis et al. 2002). The data reduction ensures that the flux ratio of 
adjacent emission lines is reliable. 

The measurement of equivalent widths has been done using a fully 
automatic procedure by fitting Gaussian profiles to both absorption and 
emission lines simultaneously (see Lewis et al. 2002 for a detailed 
discussion of the procedure and the determination of the fitting quality). 
The measured spectral features include \oii, \hbeta, \oiii, \halpha, \nii, 
and \siii. All strong emission lines needed to reliably estimate the nature 
of the main ionizing source and the gas phase oxygen abundance are measured.

Starting from the original 2dFGRS sample of 269\,013 spectra, we have applied 
different quality criteria to draw a sub-sample of 7\,402 galaxies. This 
sub-sample was selected as follows.

To get accurate estimate of the gas phase abundances and to avoid any bias 
due to observational problems, we have restricted ourselves to galaxies with 
high-quality spectra. We first exclude all galaxies observed before 
31th August 1999, which exhibit problems due to a fault of 
the atmospheric dispersion compensator within the 2dF instrument (see Lewis 
et al. 2002 for more detailed discussion on this issue). For these galaxies, 
the fitting procedure to determine line properties leads to poor quality results. 
In addition, only galaxies with good overall spectrum quality are selected.
We thus selected galaxies for which we have the best quality fit, i.e. no bad 
pixels detected within $2-3\sigma$ of the line center, for emission lines used 
to determine both the nature of the main ionizing source and the oxygen abundance, 
namely \oii, \hbeta, \oiii, \halpha, \nii, and \siii.

We also choose spectra with a median signal-to-noise ratio (measured between 
$4000-7500$\AA) of at least 10. Indeed, we found during preliminary investigations 
that adding more noisy spectra will introduce a bias in the $L-Z$ relation.
Finally, for galaxies observed several times we kept the spectrum with the 
best overall quality or, if the same, the best signal-to-noise ratio.

\subsection{Nature of the main ionizing source: starburst or AGN?}
\label{iosource}

The selection criteria applied up to now focus only on the quality of the data. 
The restricted sub-sample built so far using these criteria includes emission-line 
galaxies with different ionizing sources, i.e., either young massive stars 
related to a recent starburst or a non-thermal continuum produced by an Active 
Galactic Nucleus (AGN). As we are interested in star-forming galaxies only, 
we have to exclude galaxies for which the emission-line spectrum is typical 
of AGNs. AGN-like spectra are distinguished from starburst ones by stronger 
collisional emission lines relative to recombination lines. 

Seyfert 1 galaxies are distinguished from the other types of AGNs (Seyfert 2, LINERs) 
by broader Balmer emission lines. The distribution of Full Width at Half-Maximum 
(FWHM) for \oii, \oiiib, and \hbeta\ emission lines shows that the maximum value 
for \oii\ and \oiiib\ is 10\AA, whereas 49 galaxies have FWHM(\hbeta) exceeding 
this limit. These galaxies have been classified as Seyfert 1 galaxies, and thus 
excluded from the sample.

The widely used technique to distinguish starburst galaxies from narrow-line AGNs 
invokes diagnostic line ratios (e.g. Baldwin et al. 1981; Veilleux \& Osterbrock 1987). 
The large wavelength coverage of 2dFGRS spectra allows the easy identification of an 
AGN-dominated spectrum (Seyfert 2 and LINER) by the presence of high \nii/\halpha\ 
and \sii/\halpha\ flux ratios relative to \oiiib/\hbeta.

Figure \ref{diag} shows the location of our galaxy sample in the diagnostic 
diagrams \nii/\halpha\ vs. \oiiib/\hbeta\ and \sii/\halpha\ vs. \oiiib/\hbeta. 
We used equivalent width (EW) ratios instead of flux ratios in these diagrams.
Note that these ratios are weakly sensitive to reddening and to uncertainties 
in the spectrophotometry.
The left panel of Figure \ref{diag} shows that there are two well-separated 
sequences of emission-line galaxies in the \nii/\halpha\ vs. \oiiib/\hbeta\ diagram. 
A similar distribution has been found for a sample of galaxies drawn from the 
Sloan Digital Sky Survey (SDSS) and discussed recently by Kauffmann et al. (2003). 
Our sample contains galaxies with a large variety of excitation levels,
suggesting that this sample contains both low-metallicity and metal-rich galaxies. 
This sample is thus suitable for studying the $L-Z$ relation over a large range 
of metallicities. The AGN sequence originates from the bottom of the location of 
starburst galaxies, i.e. where metal-rich galaxies locate. This suggests that 
the inclusion of AGNs in a sample used to investigate the $L-Z$ relation will 
contaminate only the high-metallicity end. 

The location of starburt galaxies relative to AGNs in the diagnostic diagrams 
has been extensively investigated in the literature. Recently Kewley et al. (2001) 
have coupled stellar population and photo-ionization models to predict the 
distribution of starburst galaxies in the diagnostic diagrams.  
Theoretical model predictions have associated uncertainties. Several input 
parameters, such as massive star atmosphere models, stellar evolutionary tracks, 
depletion factors and the slope of the initial mass function are not known with 
high accuracy. Observationally, starburst galaxies show correlations between 
their intrinsic properties, and consequently may scatter around any separation 
in the diagnostic diagrams.  The solid line in the \sii/\halpha\ vs. \oiiib/\hbeta\ 
diagram gives an indication of the model uncertainties, and represents an upper 
limit to the theoretical boundary between starbursts and AGNs, corresponding to 
Kewley et al. (2001) model + 0.15 dex.

In both diagnostic diagrams, the predicted separation of Kewley al. (2001) 
between star-forming galaxies and AGNs match quite well the observed separation 
between starbursts and AGNs. We also show, in the \nii/\halpha\ vs. \oiiib/\hbeta\ 
diagram, the separation limit adopted by Kauffmann et al. (2003) for SDSS galaxies. 
Unfortunately, they do not show either the location of SDSS galaxies in the 
\sii/\halpha\ vs. \oiiib/\hbeta\ diagram or the separation between starbursts
and AGNs in this diagram. We found the separation limit adopted by Kauffmann et al. 
(2003) quite strict as it increases the number of galaxies with inconsistent 
classifications. We thus decided to adopt the theoretical seperation of Kewley et al.
(2001). 

\begin{figure}
\includegraphics[clip=,width=0.45\textwidth]{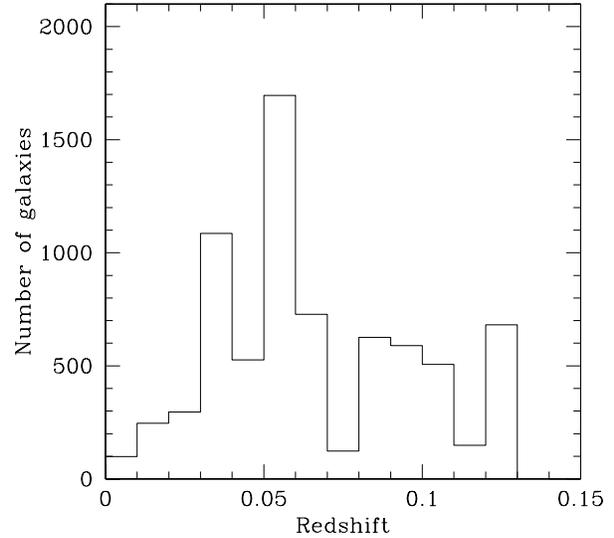}
\caption{Redshift distribution of 7\,085 star-forming galaxies selected 
from the 2dFGRS sample. The mean value is $0.05$.}
\label{histoz}
\end{figure}

We classified galaxies as starburst or AGN only if the classification criteria 
using both diagnostic diagrams are consistent, and reject all galaxies with a different 
classification in each diagnostic diagram. We end up with 7\,085 starburst galaxies, 
133 narrow-line AGNs (113 Seyfert 2 and 20 LINERs) and 135 galaxies with an ambiguous 
classification.
The completeness of the final spectroscopicaly-selected sample is difficult to define.
By restricting the sample to the presence of all the emission lines, from \oii\ 
to \sii, used both to classify the galaxies (starburst/AGN) and 
to calculate the gas-phase oxygen abundances, we implicitly impose an upper 
redshift limit of $z\sim 0.2$. By also requiring high-quality data, we exclude all 
galaxies with redshifts higher than $z>0.15$. Our sub-sample thus remains incomplete 
at higher redshift and fainter magnitudes. Since the analysis of the $L-Z$ relation 
does not rely strongly on the completeness level of the galaxy sample, and in order 
to use the largest number of galaxies for the analysis, we have decided to keep all 
the galaxies that satisfy our selection criteria listed above. Figure \ref{histoz}
shows the redshift distribution of the 7\,085 star-forming emission-line galaxies 
we will use to derive the local $L-Z$ relation. The median redshift of this sample 
is $<z>=0.05$. 

Taking advantage of the large sample of emission-line galaxies from 2dFGRS, we 
investigate the location of starbursts and narrow-line AGNs (Sey 2) in 
diagnostic diagrams involving ``blue'' emission lines only, i.e. \oii, \hbeta, and 
\oiiib. Such diagrams could be very useful to discriminate AGNs from starbursts in 
spectra of intermediate-redshift galaxies obtained in large ongoing or future deep 
spectroscopic surveys (VVDS, DEEP, etc). 
Indeed, the ``red'' emission lines, such as \halpha, \nii, and 
\sii, usually used to identify AGNs in the local universe are not observable 
in the optical range (up to $\sim 9000$ \AA) for redshifts greater than $\sim 0.3$. 
Such investigations have already been performed in the past (e.g. Rola et al. 1997; 
Dessauges-Zavadsky et al. 2000) but on limited-size samples. 
  
\begin{figure*}
\includegraphics[clip=,width=0.45\textwidth]{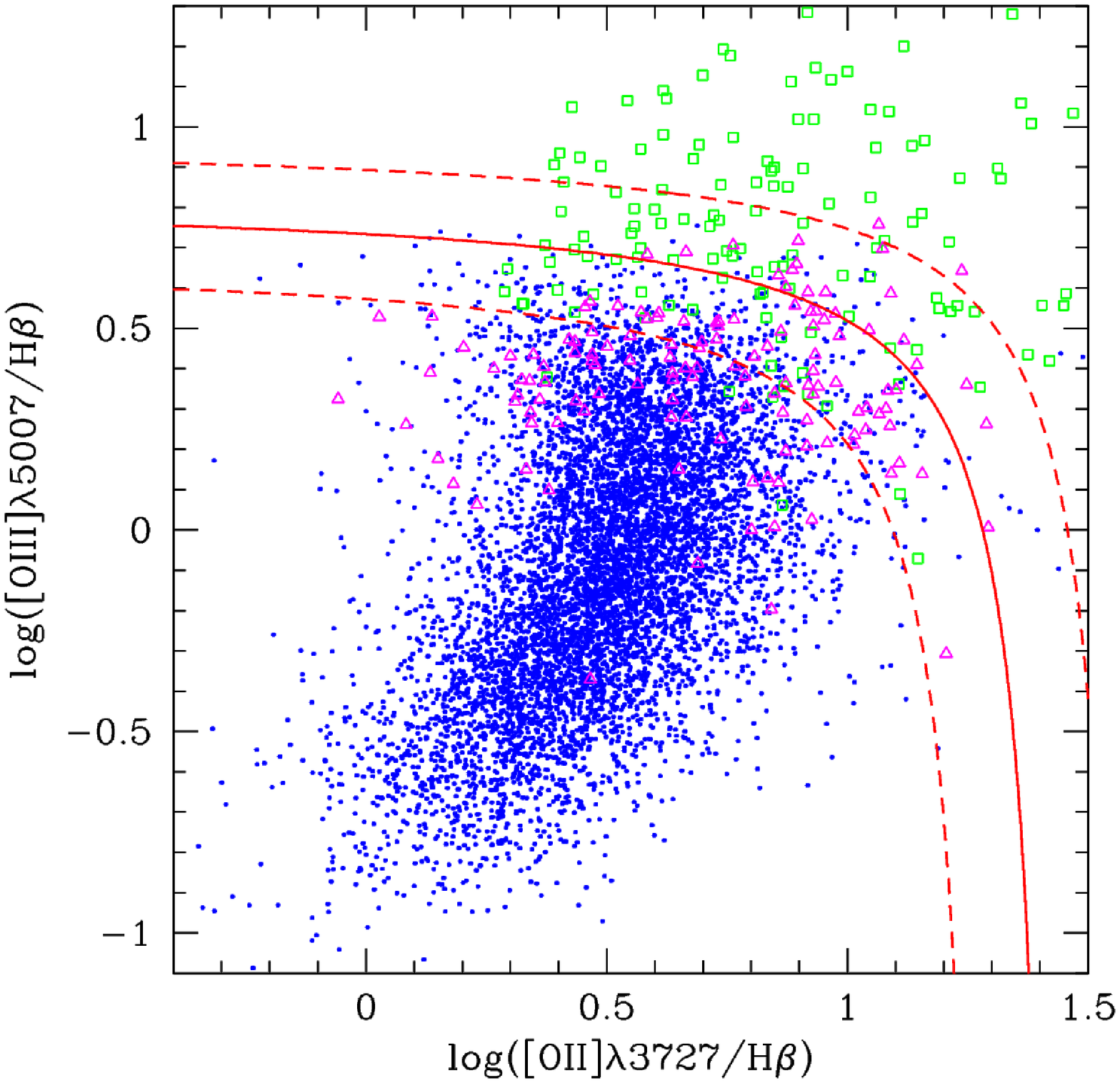}
\includegraphics[clip=,width=0.45\textwidth]{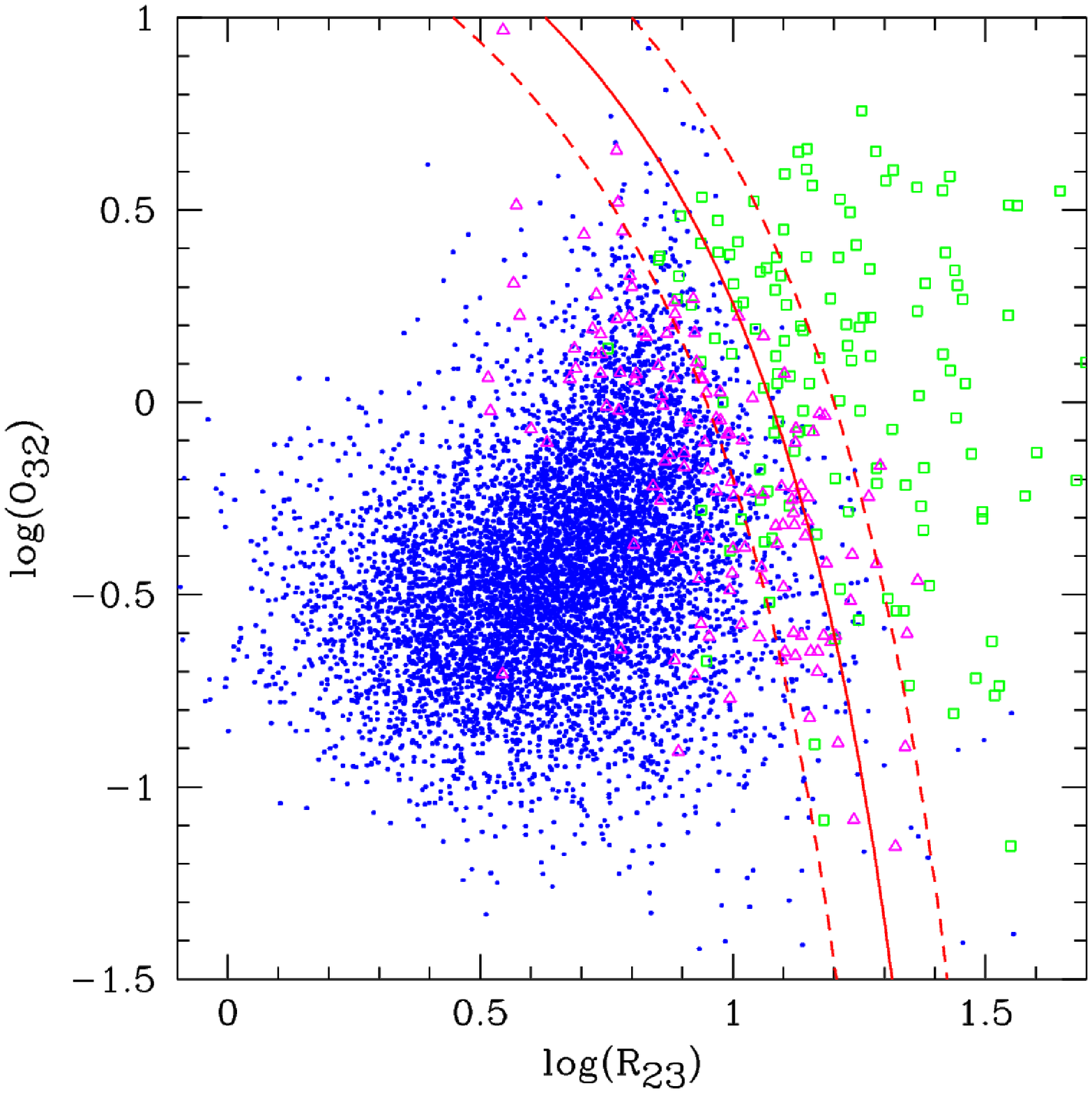}
\caption{New diagnostic diagrams for our sub-sample of 7\,353 narrow emission-line 
galaxies extracted from 2dFGRS. The continuous lines show the proposed separation between starburst 
galaxies and AGNs (see text for analytical expressions). The dashed lines show
the predicted separations plus an uncertainty of $\pm 0.15$ dex and $\pm 0.10$
in the \oiiib/\hbeta\ vs. \oii/\hbeta\ and \rr23\ vs. \oo32\ diagrams respectively.
Dots represent starburst galaxies, squares show AGNs, while triangles represent 
``unclassified'' galaxies (see text for details). These new diagrams are very 
efficient to separate AGNs from starbursts in intermediate-redshift ($z \geq 0.3$)
 galaxies for which ``blue'' emission lines only are available.}
\label{diag2}
\end{figure*}

The location of the ``classified'' narrow emission-line 2dFGRS galaxies in new 
diagnostic diagrams is shown in Figure~\ref{diag2}. The first diagram between 
\oiiib/\hbeta\ and \oii/\hbeta, proposed by Rola et al. (1997), shows a clear distinction 
between starbursts (bottom left) and AGNs (upper right). From this diagram, we define 
the following analytical expression for the demarcation curve between starbursts 
and AGNs:

\begin{equation}
\log(\frac{\rm [OIII]\lambda5007}{\rm H\beta})=\frac{0.14}{\log(\rm [OII]\lambda3727/H\beta)-1.45}+0.83
\end{equation}

Starburst galaxies are located below this line. The contamination by AGNs in this 
region is very low ($\sim 0.1$\%) taking into account $\pm 0.15$ dex uncertainties 
in the separation curve.

The second diagram between \rr23\ and \oo32\ (see Sect.~\ref{meta} for a definition of 
these ratios) shows also a clear separation between starbursts (left part) and AGNs 
(right part). From this diagram, we define 
the following analytical expression for the demarcation curve between starbursts 
and AGNs:

\begin{equation}
\log({\rm O_{\rm 32}})=\frac{1.5}{\log({\rm R_{\rm 23}})-1.7}+2.4
\end{equation}

Starburst galaxies are located to the left of this line. The discrimation between 
starbursts and AGNs is even better using this diagram, as the contamination by AGNs 
is very low ($\sim 0.1$\%) taking into account only $\pm 0.10$ dex uncertainties 
in the separation curve.

We conclude that these two diagnostic diagrams could be very efficient to separate 
AGNs from starbursts in intermediate-redshift ($z \geq 0.3$) galaxies for which 
``blue'' emission lines only are available.

\subsection{Underlying Balmer absorption}

Balmer emission-lines (e.g. \halpha, \hbeta) are generally affected by absorption 
lines from the underlying stellar populations, older than the current starburst, 
and have to be corrected. Usually the effect of the underlying stellar \hbeta\ in 
absorption is accounted for by applying an {\it ad hoc} correction to the rest-frame 
equivalent width of the \hbeta\ emission line independently of the galaxy 
properties. Different correction values for the stellar absorption have been used 
in the literature, ranging from 1\AA\ (Jansen et al. 2000) to 5\AA\ (Kennicutt 1992).   

\begin{figure}
\includegraphics[clip=,width=0.45\textwidth]{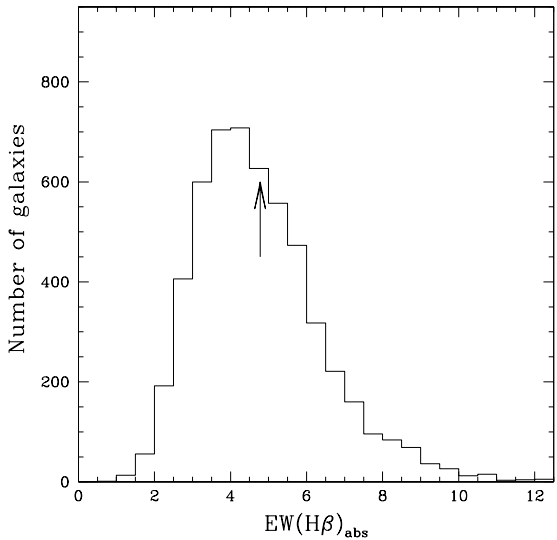}
\caption{Distribution of the EW of \hbeta\ in absorption restricted to 
galaxies with a good fit quality. The arrow indicates the mean value 
$EW({\rm H\beta)_{abs}}=4.6$\AA\ of the sample.}
\label{dist_hb}
\end{figure}

The \hbeta\ emission lines in the spectra of our galaxy sample have been corrected 
from the underlying stellar absorption by a simultaneous fit of the emission and 
absorption lines. Figure \ref{dist_hb} shows the distribution of EW$({\rm H\beta)_{abs}}$ 
restricted to galaxies with a good quality fit to \hbeta\ in absorption. The mean 
value of the EW$({\rm H\beta)_{abs}}$ is 4.6\AA, similar to Kennicutt (1992) value. 

\begin{figure}
\includegraphics[clip=,width=0.45\textwidth]{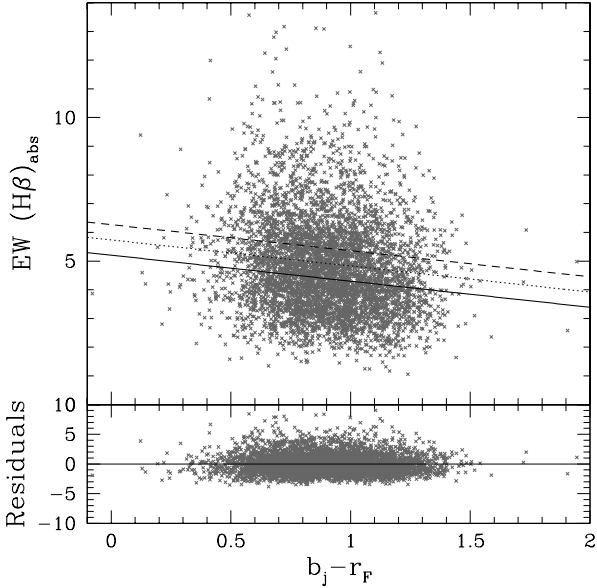}
\caption{Variation of $EW({\rm H\beta)_{abs}}$ as a function of (\textit{$b_j-r_F$}) 
color (top-panel) for our sample of starburst galaxies, restricted to galaxies with 
a good fit of \hbeta\ absorption line. Solid, dotted and dashed lines show the 
empirical calibration found for \mabs\ $= -24$, $-18$ and $-12$ respectively.
Bottom-panel shows the residuals versus the (\textit{$b_j-r_F$}) color, their rms 
is 1.75\AA.}
\label{hbabs}
\end{figure}

Instead of assuming a constant value for the EW of \hbeta\ in absorption, as usually 
done in previous studies, we tried to link the EW$({\rm H\beta)_{abs}}$ to easily 
measured galaxy parameters. We found a correlation with the (\textit{${\rm b_j-r_F}$}) 
color, an indicator of galaxy morphological type, i.e. redder is the galaxy smaller 
is the EW(${\rm H\beta)_{abs}}$. We found a lower dependancy on the absolute magnitude 
in \textit{${\rm b_j}$}-band ($M{\rm _{b_j}}$), i.e. brighter is the galaxy smaller 
is EW(${\rm H\beta)_{abs}}$. Figure \ref{hbabs} shows the EW(${\rm H\beta)_{abs}}$ 
versus (${\rm b_j-r_F}$) relation. Also shown is a linear least-squares fit to the
relation described by the following empirical calibration:
\begin{eqnarray}
{\rm EW(H\beta)_{abs}}=7.33-0.908\rmn{(b_j-r_F)}+0.0885\rmn{M_{b_j}}
\label{hb_col_mag}
\end{eqnarray}
The rms of the residuals is 1.75\AA. The EW of \halpha\ in absorption is deduced 
from EW$({\rm H\beta)_{abs}}$ using the empirical ratio 
${\rm EW(H\alpha)_{abs} = 0.75 EW(H\beta)_{abs}}$ (Gonzalez-Delgado et al. 1999). 
For a number of galaxies in our sub-sample, i.e., 1\,955, the fit for the \hbeta\ 
absorption line is of a poor quality. These spectra have been corrected using 
Eq.\,\ref{hb_col_mag}.

\subsection{Metallicity}
\label{meta}

Using estimates of the oxygen abundance as indicators of the gas-phase 
metallicity is now well documented and calibrated (e.g. Pagel 1997).
The most reliable method to derive the gas-phase oxygen abundance requires 
an estimate of the electronic temperature and density of the ionized gas 
(Osterbrock 1989). An accurate determination of these parameters requires 
reliable measurements of temperature-sensitive auroral lines, usually the 
\oiiic\ emission line. Unfortunately, the 2dFGRS spectra do not have the 
required S/N to measure correctly this line. 
The absence of direct detection of auroral lines, particularly in the case 
of metal-rich galaxies where these lines are too weak to be observed, require 
the development of alternative methods based on strong emission lines. 
The most widely used method is based on the measurement of strong emission 
lines, such as \oii, \oiii, and \hbeta. These lines contain enough information 
to get an accurate estimate of the oxygen abundance (McGaugh 1991). This is 
done through the parameter \rr23\ introduced initially by Pagel et al. (1979), 
and defined as follow:
\begin{equation}
R_{23}=\frac{{\rm [OIII]\lambda\lambda4959,5007+[OII]\lambda3727}}{{\rm H\beta}}
\end{equation}

Extensive studies have been dedicated to calibrate the relation between \rr23\ 
and oxygen abundance (McCall et al. 1985; Pilyugin 2001b). 
The small number of emission lines needed to estimate \rr23\ makes this method very 
attractive. Strong-line ratios reliably indicate the oxygen abundance to within 
the accuracy of the model calibrations, i.e. $\pm0.15$ dex.
Traditionaly, \rr23\ is estimated from the flux of emission lines. Kobulnicky 
\& Phillips (2003) have shown that the use of equivalent widths instead of fluxes 
of the same emission lines to derive \rr23\ gives almost the same results. 
The main advantage of this method is to be, at the first order, insensitive to 
the reddening. Thanks to the small wavelength separation between the emission 
lines involved in a given ratio, we are free to use the EWs of these lines 
instead of their fluxes. 

Generally, the \oiiia\ emission line has a lower signal-to-noise ratio than the 
\oiiib\ line; thus the fitting procedure of this line gives poor results 
for $\sim 40$\% of our galaxy sample. For these galaxies we have used the 
theoritical ratio \oiiib/\oiiia=2.85. The distribution of the \oiiib/\oiiia\ 
ratio, restricted to galaxies with good fitting quality of both lines, is 
in good agreement with this value, taking into account measurement uncertainties. 

A complication with the use of \rr23\ parameter to estimate the oxygen
abundance is that the dependency of metallicity on this parameter is degenerate. 
Indeed, at a fixed value of \rr23\ two different values of metallicity are possible; 
different ionization parameters should lead to similar oxygen abundances. \rr23\ 
increases with oxygen abundance in the low-metallicity regime (\doh $\leq 8.2$), 
while for metal-rich objects (\doh $\geq 8.4$) it decreases with O/H reflecting 
the efficiency of oxygen cooling over abundance in these objects. 
In the ``intermediate'' metallicity region ($8.2 <$ \doh $< 8.4$), galaxies may have 
a large range of metallicities for a tight range of \rr23. The uncertainties 
in this metallicity domain, i.e. whether an object with a given 
\rr23\ parameter lies on the metal-rich branch or on the metal-poor branch of 
the O/H vs. \rr23\ relation, dominate the uncertainties related to model 
calibrations (see above).

Different abundance indicators have been used to break the degeneracy, e.g. 
\nii/\oiiib\ (Alloin et al 1979), \nii/\oii\ (McGaugh 1994), \nii/\halpha\ 
(van Zee et al. 1998), and galaxy luminosity (Kobulnicky, Kennicutt \& Pizagno 1999). 
The first indicator is sensitive to the ionization parameter, while the galaxy
luminosity can hardly be used to break the degeneracy, first of all because 
this is exactly what we are looking for, and secondly the universality of the
$L-Z$ relation is not established yet (see Contini et al. 2002 for detailed 
discussion). To break the degeneracy between low and high metallicities, 
we used the \nii/\halpha\ secondary indicator. Galaxies with log(\nii/\halpha) 
$< -1$ are classified as low metallicity objects, while the others are assumed 
to be metal-rich galaxies. 
 
The oxygen abundance can be determined using the calibrations of
McGaugh (1991). Analytical expressions are found in Kobulnicky,
Kennicutt \& Pizagno (1999), both for the metal-poor and metal-rich regimes. 
These calibrations are parametrized as a function of \rr23\ and the ionization 
parameter defined as \oo32\ = \oiii/\oii.
We found that 495 galaxies fall into the low-metallicity region and 5\,892 
into the high-metallicity one. For 698 galaxies, the oxygen 
abundance estimated using the low-metallicity calibration was greater than 
the one derived using the high-metallicity calibration. This may happen when 
the \rr23\ parameter exceeds the maximum value allowed by the model. 
Those galaxies have been excluded from the sample in further analysis.

\section{The Luminosity -- Metallicity relation}
\label{lz}

In this section, we investigate the luminosity -- metallicity relation 
in the local universe ($z < 0.15$) using the sample of 6\,387 
star-forming galaxies extracted from the 2dFGRS as explained in 
the previous section. 

\begin{figure}
\includegraphics[clip=,width=0.45\textwidth]{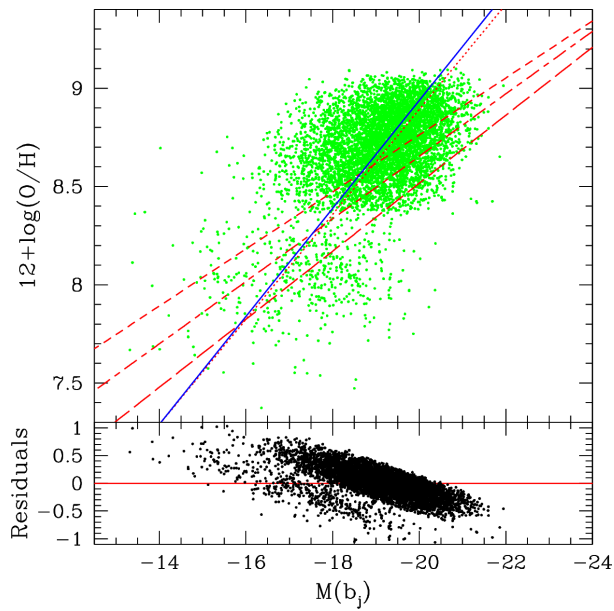}
\caption{Luminosity -- Metallicity relation for the sample of 6\,387 
star-forming galaxies extracted from the 2dFGRS. The linear regression on this 
sample is plotted as a solid line. The dotted line 
shows the relation from Melbourne \& Salzer (2002), the short-dashed line the relation 
for normal galaxies (Kobulnicky, Kennicutt \& Pizagno 1999), the long-dashed line for UV-selected 
and \hii\ galaxies (Contini et al. 2002) and the short-dashed--long-dashed line the 
relation for spirals from Pilyugin, Vilchez \& Contini (2004).
The bottom-panel shows the residuals of the linear regression.
}
\label{ohmag1}
\end{figure}

Figure \ref{ohmag1} shows our $L-Z$ relation for star-forming emission-line 
galaxies. The general trend, widely discussed in the literature
and confirmed here, is an increase of metallicty with luminosity over a large
magnitude range, from ${\rm M(b_j)=-13}$ to $-22$. Both the least-square bisector
fit (Isobe et al. 1990) and the average of the forward fit, considering the 
absolute magnitude as the independent variable, and the backward fit, considering 
the metallicity as the independent variable, yield to similar results. The bisector 
linear fit\footnote{This fitting technique is more suitable in cases in which the 
intrinsic correlation being sought measurement errors are dominated by intrinsic 
scatter.} is shown as the solid line in Figure \ref{ohmag1}, and is described by 
the following relation:  

\begin{equation}
12+\log({\rm O/H})=3.45(\pm0.09)-0.274(\pm0.005)\ \rmn{M(b_j)}
\label{ohmag}
\end{equation} 

The rms scatter around the linear fit to the relationship between metallicity 
and luminosity is equal to 
0.27 dex
, greater than the error estimates for metallicity 
determination (0.15 dex). The origin of this intrinsic scatter may be due 
to differences in the star formation history, the evolutionary status of 
the current starburst, different initial mass function, etc (see e.g. Mouhcine \& 
Contini 2002). The rms we found is identical to the one derived by Melbourne \& Salzer 
(2002), who used a sample $\sim 12$ times smaller than our sample, suggesting 
that the rms value of 0.27 dex might be the ``real'' scatter of the $L-Z$ relation.
The uncertainty on the coefficients of Equ.~\ref{ohmag} is lower than the one derived 
by Melbourne \& Salzer (2002) thanks to our larger sample. The linear regression estimated after 
eliminating the points distant for more than 3$\sigma$ from the original regression 
does not show any significant difference.

It is instructive to compare our determination of the local $L-Z$ 
relation with other published determinations. In Figure \ref{ohmag1}, we plot 
the $L-Z$ relation for local ``normal'' irregular and spiral galaxies (Kobulnicky \& 
Zaritsky 1999; short-dashed line), UV-selected and \hii\ 
galaxies (Contini et al. 2002; long-dashed line), spiral galaxies (Pilyugin et al. 
2004, short/long-dashed line), and Melbourne \& Salzer 
(2002) determination using a sample of 519 emission-line galaxies from KISS project 
(dotted line). It is worth mentioning that all determinations of the 
$L-Z$ relation have been corrected to the same cosmology. 
Our new luminosity -- metallicity relation has a steeper slope than the one found 
for nearby dwarf irregulars (Skillman et al. 1989; Richer \& McCall 1995). This 
is in contrast with Kobulnicky \& Zaritsky (1999) and Garnett (2002) finding, 
who concluded that the $L-Z$ relation exhibits a uniform growth 
over both the low- (irregular galaxies) and high-metallicity (spiral galaxies) 
regimes. On the other hand, our determination is in excellent agreement with the 
Melbourne \& Salzer (2002) one. The slopes of both determinations are almost 
identical taking into account uncertainties.  

\subsection{Possible sources of systematic errors}

\subsubsection{Different methods to derive O/H}

Before going further in our discussion, it will be useful to highlight an issue 
that may be problematic. A criticism that may be addressed is the suitability
of the strong-line method to derive oxygen abundances, and to which extent 
the systematic errors inherent to this method may affect our $L-Z$ relation. 

Different authors have pointed out that the \rr23\ method involves systematic 
errors due to the failure to take into account the variety of physical conditions 
in different \hii\ regions. For instance, Kennicutt et al. (2003) suspected the 
\rr23\ method to be systematically biased toward higher value of O/H than 
the direct method, especially at the high-metallicity end (see also Stasinska 2002).
The strong-line method yields oxygen abundances that are systematically higher than 
the electronic temperature-based technique by an amount depending on the calibration
and the excitation range considered. However, it is not clear whether the direct 
electronic temperature-based method is underestimating the oxygen abundances, or 
the strong-line method is overestimating them (see Stasinska et al. 2001 for a 
discussion of the modelling of \hii\ regions). The debate of understanding if the 
discrepancies in abundance scales are due to systematic biases in the electronic 
temperature-based scale or in the \hii\ region models used to calibrate different 
strong-line vs. abundance relations, particularly at high metallicity regime, is 
not settled yet. 

The \rr23\ parameter has been calibrated and correlated with metallicity by 
measuring oxygen abundances using the direct method, i.e. 
based on estimates of eletronic density and temperature, for a large sample of 
\hii\ regions in the low-metallicity regime and using photoionization models for
the high-metallicity regime (see Garnett 2002 for a review). 
The direct method suffers from severe limitations, such as the restricted range 
of the ionization parameters and/or metallicities and the observability of 
intrinsic weak lines. This makes this method difficult to use for abundance estimates, 
especially for galaxies at cosmological distances. There are several reasons for 
questioning the accuracy of both abundance scales (see Kennicutt et al. 2003 
for a detailed discussion of different issues related to this topic).
Pilyugin (2000) has provided new calibrations to implement McGaugh (1991) 
corrections into the empirical strong-line method. This method, usually called 
the ``P-method'', is claimed to be more accurate for low-ionization \hii\ regions 
and for metal-poor nebulae than the traditional \rr23\ method. However, the 
P-method seems to suffer some weaknesses even within the abundance range for 
which it was designed for (Saviane et al. 2002; Kennicutt et al. 2003).
Edmunds \& Pagel (1984) and Pilyugin (2000) calibrations, between \rr23\ and O/H, 
have the same slope but are shifted towards higher abundances by $\sim 0.07$ dex, 
a value smaller than both the intrinsic uncertainties of the calibration and the 
intrinsic scatter of the luminosity -- metallicity relation. 
Moreover, Melbourne \& Salzer (2002) have shown that using Pilyugin (2000) 
calibration instead of Edmunds \& Pagel (1984) one has only a small effect on 
the slope of the $L-Z$ relation. 

To check any dependancy of the $L-Z$ relation on the choice of the O/H vs. \rr23\ 
calibration, we used the Pilyugin (2000) and the McGaugh (1991) calibrations 
to derive oxygen abundances. In each case, both line fluxes and equivalent widths 
are used to calculate the \rr23\ and \oo32\ parameters. We have to mention that 
these calibrations are roughly parallel in the O/H -- \rr23\ diagram; the 
McGaugh (1991) calibration being systematicaly shifted toward higher 
abundances by $\sim 0.2$ dex at a given \rr23. In each case, we performed a 
linear regression on the $L-Z$ relation, in a similar way used to derive 
Equ.~\ref{ohmag}. We find that {\it i)} the slopes of the different $L-Z$ 
relations are similar, {\it ii)} none of the adopted O/H -- \rr23\ calibration 
leads to a slope as shallow as the $L-Z$ relation slope for irregular galaxies. 
We conclude that while the slope of the luminosity -- metallicity relation is 
slightly sensitive to the choice of the O/H -- \rr23\ calibration, our determination 
is not biased toward a steep slope. 
As our first motivation is to build a luminosity -- metallicity relation for 
the local universe to be used as a reference for intermediate and high-redshift 
studies where the majority of abundance estimates are based on the strong-line 
method (e.g. Lilly et al. 2003; Kobulnicky et al. 2003; Lemoine-Busserolle 
et al. 2003; Lamareille et al., in preparation), and since it is a ``differential'' 
comparison of the abundance properties of different galaxies, the accurate choice 
of the \rr23\ calibration is not a critical issue.  

\subsubsection{Reddening}

Is there any other systematic effect that may be responsible for the steep $L-Z$  
relation we derived? The luminosity -- metallicity relation could be biased toward 
a steeper relation as a result of neglecting the effect of internal extinction. 
One may expect naively that bright galaxies are more affected by internal 
reddening than fainter ones, causing a systematic effect which increases 
as a function of the absolute magnitude and leading to a steeper $L-Z$ relation.
To investigate this effect, we have to estimate the extinction for our sample of 
galaxies. We derived the intrinsic interstellar reddening using the observed 
Balmer line flux ratios \halpha/\hbeta\ following the standard prescription 
(e.g. Osterbrock 1989), and using the extinction law from Seaton (1979). Note that 
although different extinction laws available in the literature are different in the UV, 
they show similar behaviour in the optical, making the results of the subsequent 
analysis independent of the chosen extinction law (discussion of the galactic 
internal reddening and how its behaves as a function of galaxy properties will 
be discussed in more detail in a forthcoming paper). 
 
\begin{figure}
\includegraphics[clip=,width=0.45\textwidth]{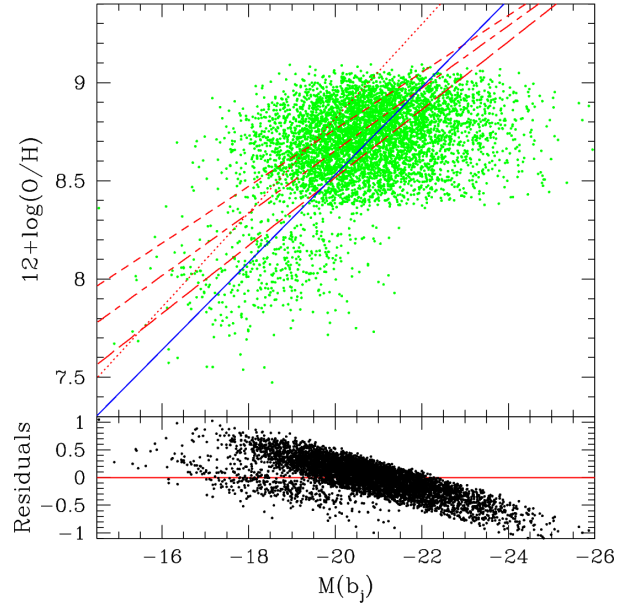}
\caption{Extinction-corrected luminosity -- metallicity relation for the sample 
of 6\,387 star-forming galaxies extracted from the 2dFGRS. The linear regression 
on this sample is plotted as a solid line. The dotted line 
shows the relation from Melbourne \& Salzer (2002), the short-dashed line the relation 
for normal galaxies (Kobulnicky, Kennicutt \& Pizagno 1999), the long-dashed line for UV-selected 
and \hii\ galaxies (Contini et al. 2002) and the short-dashed--long-dashed line the 
relation for spirals from Pilyugin, Vilchez \& Contini (2004).
The bottom-panel shows the residuals of the linear regression.
}
\label{ohmag_extcorr}
\end{figure}

The results of the extinction correction on the $L-Z$ relation, 
using the O/H -- \rr23\ calibration as given by Kobulnicky 
et al. (1999), is shown in Figure~\ref{ohmag_extcorr}. Again, a linear least 
square bisector fit is applied to the data, and we find the following expression to 
the luminosity -- metallicity relation: 

\begin{equation}
12+\log({\rm O/H})=4.07(\pm0.09)-0.223(\pm0.004)\ \rmn{M(b_j)}
\label{equationohmag}
\end{equation} 

with an rms of 
0.32. 
The extinction-corrected $L-Z$ relation is slightly 
shallower than the previous one (uncorrected for extinction), but we still 
find a much steeper slope than what was reported before in the literature. 
Interestingly enough, the slope of our extinction-corrected $L-Z$ relation 
is again in good agreement with the one found by Melbourne \& Salzer (2002) 
after correcting their photometry for the internal extinction, despite the 
different methods adopted to estimate the internal reddening correction. 
The shift in the zero point between the two relations may be due to the 
different procedures used to correct for extinction. Melbourne \& Salzer 
(2002) have corrected only galaxies that deviate at $2\sigma$ from the 
observed colour-magnitude relation, while we correct galaxies independently 
of their location in the observed colour-magnitude diagram.   

\subsection{Discussion on the slope of the $L-Z$ relation}

None of the possible sources of systematic errors listed above, i.e. 
the choice of the O/H -- \rr23\ calibration and the internal reddening, 
can explain the steep slope of the $L-Z$ relation we find. How can we explain 
the difference between our relation and the one derived for ``normal'' irregulars 
and spirals?
The fact that different determinations of the luminosity -- metallicity relation 
for irregular and spiral galaxies, using different methods 
and techniques, have similar slopes suggests that the systematic errors 
inherent to the O/H -- \rr23\ calibration cannot bias the $L-Z$ relation. 
Hence, the observed discrepancy between our relation and the other quoted 
determinations cannot be explained entirely by the uncertainties affecting 
the calibrations.  

The difference between these relations may be attribuable to the different 
samples used by various authors. We think that the nature of the galaxy sample 
used to study the luminosity -- metallicity relation has a strong impact on the 
derived slope. Mouhcine \& Contini (2002) have investigated the $L-Z$ relation for 
samples of \hii\ galaxies, UV-selected galaxies, and optical/far-infrared selected 
starburst nucleus galaxies. A linear fit to the entire sample provides a steeper 
correlation than what is found by Kobulnicky \& Zaritsky (1999), i.e. a slope of 
0.25. Once the starburst nucleus galaxies are excluded from the sample, the linear 
fit is consistent with shallower determinations of the $L-Z$ relation 
(Richer \& McCall 1985; Kobulnicky \& Zaritsky 1999; Contini et al. 2002; 
Pilyugin et al. 2004). Again, this suggests that the $L-Z$ relation is primarily sensitive 
to the nature of galaxy samples. The linear fit to irregular and spiral galaxies 
shown in Figure~\ref{ohmag1} is almost parallel to the fit to the UV-selected 
and \hii\ galaxies, i.e. at a given metallicity star-forming galaxies are brighter 
than ``normal'' galaxies. This is consistent with the expected effect of the 
current star formation episode in these galaxies, decreasing for a short period 
their mass-to-light ratio.    
As discussed by Melbourne \& Salzer (2002), different recent determinations of 
the $L-Z$ relation over a large range of metallicities/luminosities
deviate from the relation derived for dwarf galaxies, the former 
being steepper (Zaritsky et al. 1994; Pilyugin \& Ferrini 2000).  
The 2dFGRS galaxy sample contains a mixture of galaxies of different types, 
representative of the diversity of galaxies in the local universe. Thus, we believe 
that our determination of the luminosity -- metallicity relation is consistent and 
more representative of the $L-Z$ relation in the local universe than the previous 
determinations restricted to a particular type of galaxies.

\begin{figure}
\includegraphics[clip=,width=0.45\textwidth]{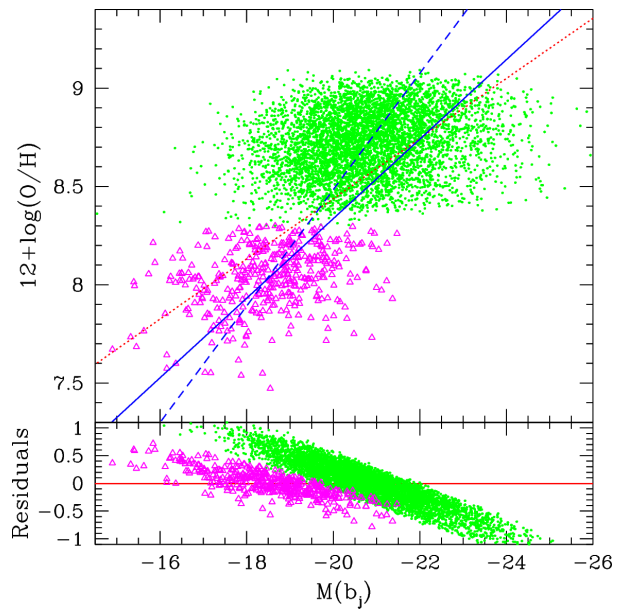}
\caption{Luminosity--metallicity relations for 2dFGRS metal-poor 
(${\rm 12+\log(O/H)\le\,8.3}$; triangles) and metal-rich 
(${\rm 12+\log(O/H)\ge\,8.3}$; dots) galaxies. The solid line shows a linear fit 
to metal-poor galaxies, dashed line shows the fit for metal-rich galaxies, 
and the dotted line shows the $L-Z$ relation of dwarf irregular galaxies 
by Skillman et al. (1989).
The bottom-panel shows the residuals of the linear regression.
}
\label{ohmag3}
\end{figure}

The difference between our $L-Z$ relation and the one derived for
dwarf galaxies suggests that the overall form of the luminosity -- metallicity relation
may not be simply approximated by a single linear relation. We split our sample of 
galaxies into two sub-samples, the first one having ${\rm 12+\log(O/H)\le\,8.3}$, as 
a representative sample of nearby metal-poor galaxies, and a second one having 
${\rm 12+\log(O/H)\ge\,8.3}$ to mimic the metal-rich galaxy population. 
Figure \ref{ohmag3} shows the $L-Z$ relation for these sub-samples. 
Also shown are {\it i)} the linear fit to the metal-poor sample (solid line) given by 
the following equation: 
${\rm 12+\log(O/H)=4.29(\pm 0.42)-0.20(\pm 0.02)\ M(b_j)}$
 compared to 
Skillman et al. (1989) $L-Z$ relation for nearby irregular 
(metal-poor) galaxies (dotted line), and {\it ii)} the linear fit to the metal-rich 
sample (dashed line) given by the following equation: 
${\rm 12+\log(O/H)=2.57(\pm 0.32)-0.30(\pm 0.02)\ M(b_j)}$.
The luminosity -- metallicity relations derived for these sub-samples are different, with 
metal-rich galaxies following a steeper $L-Z$ relation than the metal-poor galaxies. 
The slope of the dwarf galaxy $L-Z$ relation is slightly steeper than the one derived 
by Skillman et al. (1989), for which the oxygen abundances were estimated using the direct 
electronic-temperature method. Restricting the 2dFGRS sub-sample to the best quality spectra, 
we found a slope ($0.164\pm 0.02$), for the dwarf galaxy 
$L-Z$ relation, similar to what has beenderived by Skillman et al. (1989), i.e. a slope of 0.153. 
The variation 
in the derived slopes being small and included in the error bars, it shows that the choice 
of the 2dFGRS sub-sample resulting from the selection criteria does not change significantly 
the slope of the derived $L-Z$ relations. The small differences in the slope and zero 
point between our $L-Z$ determination and the one by Skillman et al. (1989) for metal-poor 
galaxies may be due to: {\it i)} the different methods applied to derive O/H 
in low-mass galaxies; Skillman et al. (1989) used the ``direct'' 
method as we used the strong-line one, {\it ii)} the filter used to measure 
the galaxy magnitude varies also; $b_{\rm j}$-filter in our case instead of 
the standard $B$-filter used by Skillman et al. (1989), and {\it iii)} the limited-size of the 
Skillman et al.'s sample compared with the 2dFGRS sub-sample.

\subsection{Dependance with other galaxy properties}
\label{otherparam}

%
%
%
%
%

The large sample presented in this paper gives an opportunity to investigate a 
possible sensitivity of the luminosity -- metallicity relation to various galaxy 
properties, such as star formation rate (SFR), galaxy stellar content, etc.  

At a given mass, starburst galaxies tend to be brighter than ``normal'' galaxies 
forming stars at a lower rate, especially in the $B$ band. This could introduce 
a systematic shift in the $L-Z$ relation for starburst galaxies toward higher 
luminosities for a given metallicity. Unfortunately, the 2dFGRS spectra are not 
calibrated in absolute flux, preventing us from estimating the current SFR of 
galaxies. Rather then using fluxes, we thus use Balmer emission-line EWs. The 
Balmer emission-line EWs are a measure of the relative proportion of ionizing 
photons (produced by massive stars related to the current star formation event) 
and continuum photons (produced by a mix between the whole cluster embedded in 
the \hii\ region, the underlying older stellar population, and a contribution 
from the ionized gas). 
The Balmer emission-line EWs may be understood as a measurement of the specific 
SFR. An analysis of the luminosity -- metallicity relation as a function of the 
EW of Balmer emission lines might therefore tell us something about the 
sensitivity of this relation to the ``current'' SFR of galaxies.    

We split the sample into different groups as a function of their ${\rm EW(H\alpha)_e}$. 
The first group contains galaxies having ${\rm EW(H\alpha)_e > 40}$\AA, group II 
contains galaxies with ${\rm 20\AA\le EW(H\alpha)_e\le 40\AA}$, group III contains
those with ${\rm 10\AA\le EW(H\alpha)_e\le 20\AA}$, and finally group IV contains
galaxies having ${\rm EW(H\alpha)_e\le 10\AA}$. We then perform, for each group, 
a linear least square bisector fit to the $L-Z$ relation. 
We find that the luminosity -- metallicity relations for the first two groups are 
similar, whithin the errors, to each other and to the whole sample $L-Z$ relation, 
showing similar scatter around the fit. This might be expected as the first two bins 
are sampling a short starburst age interval (Stasinska et al. 2001), within which 
the luminosity is not expected to vary significantly. 
The two remaining groups show $L-Z$ relations with different slopes, 
indicating a sensitivity of the latter to the current SFR. However, the scatter 
around the fits are large, preventing us from drawing any firm conclusion. 

Similarly, we investigate the sensitivy of the luminosity -- metallicity relation 
to the galaxy colour. Again, no firm conclusion can be drawn. Furthermore there is 
no obvious correlation between the residuals from the $L-Z$ relation in one hand and 
${\rm EW(H\alpha)_e}$ and colour in other hand. The present sample of galaxies does 
not present any appealing evidence for the sensitivity of the luminosity -- metallicity 
relation to the stellar content of galaxies. 

\subsection{The $L-Z$ relation in the $R$ band}

$r_{\rm F}$-band photometry is available for the 2dFGRS galaxy sample. We thus 
investigate the $r_{\rm F}$-band luminosity -- metallicity relation and find 
a trend between these two quantities similar to what is found in the ``blue'' band. 
We fit a linear relation between the oxygen abundance and the $r_{\rm F}$-band absolute 
magnitude, in a similar way to what has been done for the $b_{\rm j}$-band relation. 
We find the following $r_{\rm F}$-band $L-Z$ relation: 

\begin{equation}
12+\log({\rm O/H})=3.72(\pm0.08)-0.249(\pm0.004)\ \rmn{M(r_F)}
\label{ohmagr}
\end{equation} 

with an rms of the residuals of 0.25 dex, lower than the scatter for the 
$b_{\rm j}$-band $L-Z$ relation. This might be due to the smaller 
sensitivity of $r_{\rm F}$-band luminosity to the occurrence 
and the evolutionary status of a recent star formation episode.

\subsection{Dependance with redshift and selection biases}

\begin{figure}
\includegraphics[clip=,width=0.45\textwidth]{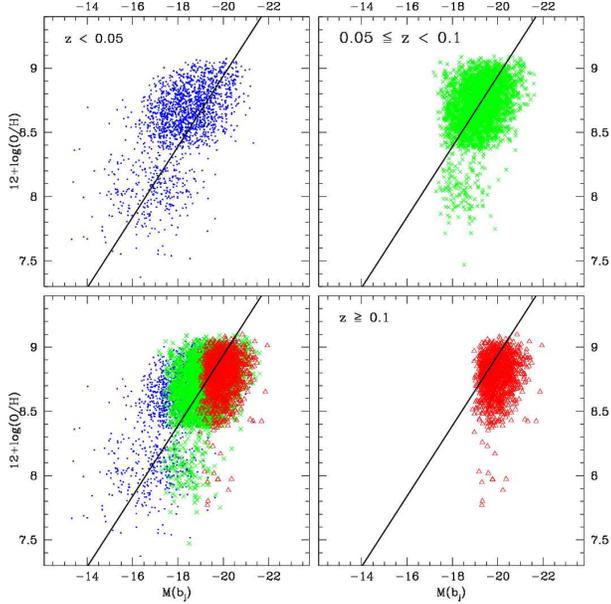}
\caption{Luminosity--Metallicity relation for different bins of redshift: 
$z<0.05$ (upper-left, blue dots),
$0.05\le{}z<0.1$ (upper-right, green crosses) and $z\ge{}0.1$ 
(bottom-right, red triangles). The bottom-left
panel show the three redshift bins together. The solid line shows the $L-Z$ 
relation given in Eq.\ref{ohmag}.}
\label{ohmag4}
\end{figure}

In this last section, we investigate whether the $L-Z$ relation 
is redshift-dependant as recently suggested in Schulte-Ladbeck et 
al. (2003) using SDSS data.
Despite the small redshift range of our sub-samble ($0<z<0.15$), 
Figure~\ref{ohmag4} shows indeed significant differences if we divide the 
sample into three different redshift bins (i.e. $z<0.05$, $0.05\le{}z<0.1$ and 
$z\ge{}0.1$). At a given metallicity, the available
observed luminosity range seems to be shifted toward higher luminosities 
when we look further in the past. This effect
is supported by calculating, for each bin, the mean value of the luminosity, 
which increases ($-18.2$ for $z<0.05$,
$-19.2$ for $0.05\le{}z<0.1$ and $-20.0$ for $z\ge{}0.1$) while, in average,  
the metallicity remains approximatively constant (\doh\ = 8.58, 8.70 and 8.77 respectively).
This effect is particulary strong for low-metallicity objects 
(\doh\ $< 8.3$, see the bottom-left panel).

Before drawing any conclusion on the evolution of the $L-Z$ relation with redshift, 
one must take into account some selection effects. First of all, we have to 
consider that high-luminosity objects will preferably appear at high redshifts, 
where the volume of the observed universe is large enough to make their number 
significative given their small intrinsic proportion. 
In the opposite way, low-luminosity objects may appear at all redshifts, 
but most of them are observed at low redshift due to the magnitude-limited nature 
of the survey (the Malmquist bias). 
The first effect does not explain the evolution of the mean luminosity, 
while it is normalized to the total number of
observed objects. The last one is more problematic because we loose 
almost all low-metallicity objects at higher redshift.
Our quality selection criteria also introduce a bias: very 
metal rich, and thus very luminous, objects have been rejected from 
our sub-sample by the quality criterium on oxygen 
emission lines, considering that they are very faint (low S/N) for 
\doh\ $> 9$.

Taking into account these selection effects, no firm conclusion 
can be drawn from this study on any evolution of the $L-Z$ relation 
with redshift.

\section{Conclusions} 
\label{concl}

We have investigated the luminosity -- metallicity relation in the local universe, 
using data of $\sim 7000$ star-forming galaxies ($0 < z < 0.15$) extracted from the 
2dFGRS spectroscopic data set. The sample used in this paper is by far the largest 
sample of galaxies to date to derive the $L-Z$ relation.   

We first distinguished star-forming galaxies from AGNs using ``standard'' diagnostic 
diagrams to build a homogeneous sample of 6\,387 starburst galaxies for the $L-Z$ 
study. We propose also new diagnostic diagrams using ``blue'' emission lines (\oii, 
\oiiib, and \hbeta) to discriminate starbursts from AGNs in intermediate-redshift 
($z > 0.3$) galaxies. We have shown that the underlying absorption lines affecting 
Balmer lines in emission depend on galaxy properties, with a wide distribution ranging
from 1 \AA\ to 10 \AA\ with an average value of 4.6\AA. Oxygen-to-hydrogen abundance ratios 
were estimated for this sample of star-forming galaxies using the strong-line method, 
which relates the \rr23\ and \oo32\ parameters to O/H. We confirmed the existence of 
the luminosity -- metallicity relation over a large range of abundances and luminosities. 
We find a linear relation between the gas-phase oxygen abundance and both the ``raw'' 
and extinction-corrected $b_{\rm j}$-band absolute magnitude with a rms of $\sim 0.26$. 
A similar relation, with nearly the same scatter, exists for the $r_{\rm F}$-band absolute 
magnitude. 

Our determination of the $L-Z$ relation is in a remarkably good agreement with the 
one derived by Melbourne \& Salzer (2002) using the KISS data . However, the slope 
of the relation we derived is much steeper than the ones previously determined 
using different samples of ``normal'' dwarf and spiral galaxies. 
We argue that this difference is not due to any systematic error inherent to the 
method used to derive O/H in galaxies. The nature of the galaxy 
sample used to investigate the $L-Z$ correlation is however crucial. Neglecting 
a certain type of galaxies, such as starburst nucleus galaxies, in previous studies has 
biased the determination of the $L-Z$ relation toward shallower 
slopes. The $L-Z$ relation restricted to metal-poor galaxies of our sample is in
agreement with the relation derived by Skillman et al. (1989) for the low-metallicity 
irregular galaxies.

The luminosity -- metallicity relation extends over a range of $9$ magnitudes in 
luminosity, and a factor of $\sim 50$ in oxygen abundance. The 2dFGRS sample of 
galaxies used in this paper contains a large diversity of masses and stellar 
populations, from dwarf galaxies to massive spirals. By using such a sample of 
galaxies, we are constructing a more general luminosity -- metallicity relation 
in the local universe than previous studies restricted to a given class of galaxies.   
This correlation seems to be continuous from faint dwarf galaxies to massive spirals, 
implying that the physical mechanism(s) regulating this correlation is common to 
star-forming galaxies over the whole Hubble sequence, without a clear dependency on 
the stellar content of galaxies.

The investigation of the chemical properties of high-$z$ galaxies, and the comparison 
between galaxy properties over a large range of redshifts will certainly help to 
clarify the evolutionary pattern of galaxies at different cosmic epochs. 

\section*{Acknowledgments}
M. M. would like to thank warmly A. Lan\c{c}on for interesting and highlighting 
discussions about reddening in galaxies. We are grateful to the anonymous referee 
for useful comments and suggestions.

\label{lastpage}

\begin{thebibliography}{}
\bibitem[\protect\citeauthoryear{Alloin et al.}{1979}]{alloin79} 
Alloin D., Collin-Souffrin S., Joly M., Vigroux L., 1979, A\&A, 78, 200 
\bibitem[\protect\citeauthoryear{Baldwin et al.}{1981}]{baldwin81} 
Baldwin J.~A., Phillips M.~M., Terlevich R., 1981, PASP, 93, 5 
\bibitem[\protect\citeauthoryear{Brodie \& Huchra}{1991}]{brodie91}
Brodie J., Huchra J.P., 1991, ApJ, 379, 157
\bibitem[\protect\citeauthoryear{Colless et al.}{2001}]{colless01}
Colless M., Dalton G., Maddox S.J., et al., 2001, MNRAS, 328, 1039 
\bibitem[\protect\citeauthoryear{Contini et al.}{2002}]{contini02}
Contini T., Treyer M. A., Sullivan M., Ellis R. S., 2002, MNRAS, 330, 75
\bibitem[\protect\citeauthoryear{Dessauges-Zavadsky et al.}{2000}]{dessauges00}
Dessauges-Zavadsky M., Pindao M., Maeder A., Kunth D., 2000, A\&A, 355, 89 
\bibitem[\protect\citeauthoryear{Edmunds \& Pagel}{1984}]{edmunds84}
Edmunds M.~G., Pagel B.~E.~J., 1984, MNRAS, 211, 507 
\bibitem[\protect\citeauthoryear{Erb et al.}{2003}]{erb03}
Erb D.K., Shapely A.E., Steidel C.C., et al., 2003, ApJ, 591, 101
\bibitem[\protect\citeauthoryear{Garnett \& Shield}{1987}]{garnett87}
Garnett D.R., Shield G.A., 1987, ApJ, 317, 82
\bibitem[\protect\citeauthoryear{Garnett et al.}{1997}]{garnett97}
Garnett D.R., Shield G.A., Skillman E.D., et al., 1997, ApJ, 489, 63
\bibitem[\protect\citeauthoryear{Garnett}{2002}]{garnett02}
Garnett D.R., ApJ, 2002, 581, 1019
\bibitem[\protect\citeauthoryear{Gonzalez-Delgado}{1999}]{gonzalez99}
Gonz{\' a}lez Delgado R.M., Leitherer C., Heckman T.M., 1999, ApJS, 125, 489 
\bibitem[\protect\citeauthoryear{Hammer et al.}{2001}]{hammer01}
Hammer F., Gruel N., Thuan T.X., Flores H., Infante L., 2001, ApJ, 550, 570 
\bibitem[\protect\citeauthoryear{Im et al.}{2002}]{im02}
Im M., Simard L., Faber S.M., et al., 2002, ApJ, 571, 136
\bibitem[\protect\citeauthoryear{Isobe et al.}{1990}]{isobe90}
Isobe T., Feigelson E.~D., Akritas M.~G., Babu G.~J., 1990, ApJ, 364, 104 
\bibitem[\protect\citeauthoryear{Jansen et al.}{2000}]{jansen00}
Jansen R.A., Fbricant D., Franx M., Caldwell N., 2000, ApJS, 126, 331  
\bibitem[\protect\citeauthoryear{Kauffmann et al.}{1993}]{kauffmann93}
Kauffmann G., White S.D.M.,  Guiderdoni B., 1993, MNRAS, 264, 201
\bibitem[\protect\citeauthoryear{Kauffmann et al.}{2003}]{kauffmann03}
Kauffmann G. et al., 2003, MNRAS, 346, 1055
\bibitem[\protect\citeauthoryear{Kennicutt}{1992}]{kennicutt92}
Kennicutt R.C., 1992, ApJ, 388, 310
\bibitem[\protect\citeauthoryear{Kennicutt et al.}{2003}]{kennicutt03}
Kennicutt R.C., Bresolin F., Garnett D.R., 2003, ApJ, 591, 801
\bibitem[\protect\citeauthoryear{Kewley et al.}{2001}]{kewley01}
Kewley L. J., Heisler C. A., Dopita M. A., Lumsden S., 2001, ApJS, 132, 37
\bibitem[\protect\citeauthoryear{Kobulnicky et al.}{1999}]{kobulnicky99}
Kobulnicky H. A., Kennicutt Jr. R. C., Pizagno J. L., 1999, ApJ, 514, 544
\bibitem[\protect\citeauthoryear{Kobulnicky \& Phillips}{2003}]{kobulnicky03}
Kobulnicky H. A., Phillips A. C., 2003, ApJ, 599, 1031
\bibitem[\protect\citeauthoryear{Kobulnicky et al.}{2003}]{kobulnickyetal03}
Kobulnicky H. A., Willmer C.N.A, Weiner B.J., Koo D.C., Phillips A. C., Faber S.M., Sarajedini V.L., Simard L., Vogt N.P., 2003, ApJ, 599, 1006
\bibitem[\protect\citeauthoryear{Kobulnicky \& Zaritsky}{1999}]{kobulnicky99b}
Kobulnicky H. A., Zaritsky D., 1999, ApJ, 511, 118
\bibitem[\protect\citeauthoryear{Kobulnicky \& Koo}{2000}]{kobulnicky00}
Kobulnicky H. A., Koo D.C., 2000, ApJ, 545, 712 
\bibitem[\protect\citeauthoryear{Lemoine-Busserolle et al.}{2003}]{lemoine03}
Lemoine-Busserolle M., Contini T., Pello R., Le Borgne J.-F., Kneib J.-P., Lidman C.,  
2003, A\&A, 397, 839
\bibitem[\protect\citeauthoryear{Lequeux et al.}{1979}]{lequeux79}
Lequeux J., Peimbert M., Rayo J.F., et al., 1979, A\&A, 80, 155
\bibitem[\protect\citeauthoryear{Lewis et al.}{2001}]{lewis01}
Lewis I., Balogh M., De Propis R., et al. 2002, MNRAS, 334, 673 
\bibitem[\protect\citeauthoryear{Lilly et al.}{2003}]{lilly03}
Lilly S.J., Carrollo C.M., Stockton A.N., 2003, ApJ, 597, 730
\bibitem[\protect\citeauthoryear{Madau et al.}{1996}]{madau96}
Madau P., Ferguson H.C., Dickinson M.E., Gaivalisco M., Steidel C.C., Fruchter A.,
1996, MNRAS, 283, 1388
\bibitem[\protect\citeauthoryear{Mateo}{1998}]{mateo98} 
Mateo M., 1998, ARA\&A, 36, 435
\bibitem[\protect\citeauthoryear{McCall et al.}{1985}]{mccall85}
McCall M.L., Rybski P.M., Shields G.A., 1985, ApJS, 57, 1
\bibitem[\protect\citeauthoryear{McGaugh}{1991}]{mcgaugh91}
McGaugh S. S., 1991, ApJ, 380, 140
\bibitem[\protect\citeauthoryear{McGaugh}{1994}]{mcgaugh94}
McGaugh S.~S., 1994, ApJ, 426, 135 
\bibitem[\protect\citeauthoryear{Mehlert et al.}{2002}]{mehlert02}
Mehlert D., Noll S., Appenzeller I., et al., 2002, A\&A, 393, 809 
\bibitem[\protect\citeauthoryear{Melbourne \& Salzer}{2002}]{melbourne02}
Melbourne J., Salzer J. J., 2002, AJ, 123, 2302
\bibitem[\protect\citeauthoryear{Milvang-Jensen et al.}{2003}]{milvang03}
Milvang-Jensen B., Aragon-Salamanca A., Hau G., Jorgensen I., Hjorth J., 2003, 
MNRAS, 339, 1 
\bibitem[\protect\citeauthoryear{Mouhcine \& Contini}{2002}]{mouhcine02}
Mouhcine M., Contini T., 2002, A\&A, 389, 106
\bibitem[\protect\citeauthoryear{Osterbrock}{1989}]{osterbrock89}
Osterbrock D.E., 1989, Astrophysics of Gaseous Nebulae and Active Galactic Nuclei
                     (Mill Valley: Unvi. Sci.)
\bibitem[\protect\citeauthoryear{Pagel et al.}{1979}]{pagel79}
Pagel B.E.J., Edmunds M.G., Blackwell D.E., Chum M.S., Smith G., 1979, MNRAS, 189, 95    
\bibitem[\protect\citeauthoryear{Pagel}{1997}]{pagel97}
Pagel B.E.J., 1997, Nucleosynthesis and Chemical Evolution of Galaxies (Cambridge:
Cambridge University Press)
\bibitem[\protect\citeauthoryear{Pettini}{2003}]{pettini03}
Pettini M., 2003, to appear in ``Chosmochemistry: The Melting Pot of Elements'' (astro-ph/0303272)
\bibitem[\protect\citeauthoryear{Pettini et al.}{1998}]{pettini98}
Pettini M., Kellogg M., Steidel C.C., Dickinson M., Adelberger K.L., Giavalisco M., 
1998, ApJ, 508, 539 
\bibitem[\protect\citeauthoryear{Pettini et al.}{2001}]{pettini01}
Pettini M., Shapley A.~E., Steidel C.~C., Cuby J., Dickinson M., Moorwood A.~F.~M., Adelberger K.~L., Giavalisco M., 2001, ApJ, 554, 981 
\bibitem[\protect\citeauthoryear{Pilyugin}{2000}]{pilyugi00}
Pilyugin L.~S., 2000, A\&A, 362, 325 
\bibitem[\protect\citeauthoryear{Pilyugin \& Ferrini}{2000}]{pilyugin00}
Pilyugin L. S., Ferrini F., 2000, A\&A, 358, 72
\bibitem[\protect\citeauthoryear{Pilyugin}{2001a}]{pilyugin01a}
Pilyugin L. S., 2001a, A\&A, 374, 412
\bibitem[\protect\citeauthoryear{Pilyugin}{2001b}]{pilyugin01b}
Pilyugin L. S., 2001b, A\&A, 369, 594
\bibitem[\protect\citeauthoryear{Pilyugin et al.}{2004}]{pilyugin04}
Pilyugin L. S., V\'ilchez J. M., Contini T., 2004, A\&A, submitted
\bibitem[\protect\citeauthoryear{Richer \& McCall}{1995}]{richer95}
Richer M.G., McCall M.L., 1995, ApJ, 445, 642
\bibitem[\protect\citeauthoryear{Rola et al.}{1997}]{rola97}
Rola C.~S., Terlevich E., Terlevich R.~J., 1997, MNRAS, 289, 419 
\bibitem[\protect\citeauthoryear{Saviane et al.}{2002}]{saviane02}
Saviane I., Rizzi L., Held E.V., Bresolin F., Momany Y., 2002, A\&A, 390, 59
\bibitem[\protect\citeauthoryear{Schulte-Ladbeck et al.}{2003}]{schulte03}
Schulte-Ladbeck, R.E., Miller, C.J., Hopp, U., Hopkins, A., Nichol, R.C., Voges, W., Fang, T., 2003, to appear in ``Multiwavelength Mapping of Galaxy Formation and Evolution'' (astro-ph/0312069)
\bibitem[\protect\citeauthoryear{Seaton}{1979}]{seaton79}
Seaton M.~J., 1979, MNRAS, 187, 73P 
\bibitem[\protect\citeauthoryear{Skillman et al.}{1989}]{skillman89}
Skillman E.D., Kennicutt R.C., Hodge P.W., 1989, ApJ, 347, 875
\bibitem[\protect\citeauthoryear{Somerville \& Primack}{1999}]{somerville99}
Somerville R.S., Primack J.R., 1999, MNRAS, 310, 1087
\bibitem[\protect\citeauthoryear{Stasinska}{2002}]{stasinska02}
Stasinska G., 2002, Revista Mexicana de Astronomia y Astrofisica Conference Series, 12, 62
\bibitem[\protect\citeauthoryear{Stasinska et al.}{2001}]{stasinska01}
Stasinska G., Schaerer D., Leitherer C., 2001, A\&A, 370, 1 
\bibitem[\protect\citeauthoryear{van Dokkum \& Ellis}{2003}]{vandokkum98}
van Dokkum P.G., Ellis R.S., 2003, ApJ, 592, 53
\bibitem[\protect\citeauthoryear{van Zee et al.}{1998}]{vanzee98}
van Zee L., Salzer J. J., Haynes M. P., O'Donoghue A. A., Balonek T. J., 1998, AJ, 116, 2805
\bibitem[\protect\citeauthoryear{Veilleux \& Osterbrock}{1987}]{veilleux87}
Veilleux S., Osterbrock D. E., 1987, ApJS, 63, 295
\bibitem[\protect\citeauthoryear{Vilchez}{1995}]{vilchez95} 
Vilchez J.M., 1995, AJ, 110, 1090
\bibitem[\protect\citeauthoryear{Zaritsky et al.}{1994}]{zaritsky94}
Zaritsky D., Kennicutt R.C., Huchra J.P, 1994, ApJ, 420, 87
\bibitem[\protect\citeauthoryear{Ziegler et al.}{2002}]{ziegler02}
Ziegler B.L., Bohm A., Fricke, et al., 2002, ApJ, 564, 69
\end{thebibliography}
\end{document}